\documentclass[aps,prb,longbibliography,twocolumn,floatfix,superscriptaddress,nofootinbib]{revtex4-2}

\usepackage{graphicx}
\usepackage{amsmath}
\usepackage{braket}
\usepackage{amssymb}
\usepackage{mathtools}
\usepackage{makeidx}
\usepackage{amsfonts}
\usepackage{bm}
\usepackage{bbm}
\usepackage{xcolor} 
\usepackage{tikz}
\usepackage{multirow}
\usepackage{tikz-3dplot}
\usetikzlibrary{3d,shapes.arrows,decorations.pathreplacing, arrows.meta, positioning, matrix}
\usepackage[colorlinks=true, linkcolor=blue, citecolor=blue, urlcolor=blue]{hyperref}

\begin{document}

\title{Dephasing‑driven suppression of superradiance and metastable dynamics in the anisotropic open Rabi model}

\author{Jivyanshu Priya}
\email{jivyanshu23@iiserb.ac.in}
\affiliation{Department of Physics, Indian Institute of Science Education and Research, Bhopal, Madhya Pradesh 462066, India}
\author{Pragna Das}
\affiliation{Department of Theoretical Physics, J. Stefan Institute, SI-1000 Ljubljana, Slovenia}

\author{Auditya Sharma}
 
\affiliation{Department of Physics, Indian Institute of Science Education and Research, Bhopal, Madhya Pradesh 462066, India}
\date{\today}

\begin{abstract}
Combining phase-space tomography with Liouvillian gap scaling, we show that atomic dephasing reshapes dissipative criticality. We study the anisotropic open Rabi model in the presence of cavity photon loss, atomic relaxation, and atomic dephasing. Our stability analysis shows that adding atomic dephasing to cavity photon loss destabilizes the superradiant phase. Furthermore, we show that when atomic relaxation stabilizes a long-lived metastable superradiant state, atomic dephasing competes with this effect by reducing its coherence and shortening its lifetime. Our results demonstrate that different environmental channels can qualitatively
alter the nonequilibrium critical behavior of finite-component light–matter systems, depending not only on their strength but also on their microscopic origin. 
\end{abstract}

\maketitle
\section{Introduction}
\label{sec:introduction}
The control and engineering of open
quantum many-body systems has emerged as a central
theme in modern quantum physics, offering a route to
stabilize exotic nonequilibrium phases of matter that
have no equilibrium counterparts~\cite{diehl2008quantum, verstraete2009quantum, daley2014quantum, sieberer2016keldysh, kirton2019introduction, rossini2021coherent, harrington2022engineered, fazio2025many}. The competition between coherent interactions, external driving, and dissipation can stabilize steady states that undergo dissipative phase transitions (DPTs), characterized by nonanalytic changes in steady-state properties as external control parameters are varied~\cite{diehl2008quantum, kessler2012dissipative, sieberer2016keldysh}. Accompanying these transitions is a rich landscape of critical phenomena, including spontaneous symmetry breaking~\cite{kirton2017suppressing, soriente2018dissipation}, critical slowing down associated with the closing of the Liouvillian spectral gap~\cite{fink2018signatures, minganti2018spectral, mori2020resolving, haga2021liouvillian, mori2023symmetrized}, and long-lived metastability arising from the separation of relaxation time scales~\cite{rose2016metastability, le2017metastability, macieszczak2021theory, jin2024theory, xiao2026metastability}. These phenomena have become experimentally accessible across a wide range of platforms, including cavity- and circuit-QED systems, trapped ions, and ultracold atomic gases, stimulating extensive theoretical and experimental efforts to understand nonequilibrium quantum criticality~\cite{baumann2010dicke, brennecke2013real, schmidt2013circuit, raftery2014observation, fitzpatrick2017observation, li2024spin}.

Among the various platforms for exploring nonequilibrium criticality, light–matter systems described by the Dicke model and its few-body analog, the quantum Rabi model, have proven paradigmatic ~\cite{dicke1954coherence, das2024dicke, hwang2015quantum, wang2018quantum}. The Dicke and Rabi models describe the interaction between atoms modeled as two-level systems (TLS) and light modeled as a bosonic mode. The Dicke model exhibits a well-known normal-to-superradiant phase transition in the thermodynamic limit ($N_{\text{atoms}} \xrightarrow{}\infty)$~\cite{emary2003quantum, kirton2019introduction}, and the Rabi model shows analogous critical behavior as the qubit–cavity frequency ratio becomes large ($\Omega / \omega \xrightarrow{}\infty)$~\cite{hwang2015quantum, hwang2018dissipative}. Crucially even a single atom Rabi model undergoing cavity photon loss can exhibit a second-order dissipative phase transition~\cite{hwang2018dissipative}, establishing the open Rabi model as the minimal light–matter system with genuine quantum criticality. The anisotropic extensions of the Dicke~\cite{baksic2014controlling, buijsman2017nonergodicity, hu2021out, das2023phase, das2023periodically} and Rabi~\cite{xie2014anisotropic, liu2017universal, wang2018quantum, chen2021multiple} models, in which the rotating- and counter-rotating-wave interactions are independently tunable, have revealed an even richer phase structure. Using the stability analysis, it is shown that besides the normal phase (NP) and the superradiant phase (SP), anisotropy gives rise to a bistable phase (BP) in which both NP and SP phases are simultaneously stable as steady states ~\cite{soriente2018dissipation, liu2017universal, lin2022dissipation, lyu2024multicritical, li2024spin, wu2024experimental}. This establishes the anisotropic open Rabi model as a versatile platform for investigating nonequilibrium critical phenomena beyond the isotropic limit.

In addition to the steady-state phase diagram, dissipative phase transitions are also characterized by their relaxation dynamics, encoded in the spectrum of the Liouvillian superoperator~\cite{minganti2018spectral}. When the eigenvalues $\lambda_k$ of $\mathcal{L}$ are ordered by decreasing real part, $\lambda_0=0$ corresponds to the steady state, and each nonzero eigenvalue defines a decay rate $\Gamma_k=-\mathrm{Re}\,\lambda_k$. The asymptotic decay rate toward the steady state is set by the Liouvillian spectral gap $\Gamma_1$, the decay rate of the slowest-decaying nonzero mode. Its closing as a system parameter approaches the critical limit signals critical slowing down~\cite{minganti2018spectral, casteels2017}, while its closing in the thermodynamic limit deep inside a symmetry-broken phase signals that the symmetry-broken states are strictly stable. A gap that remains finite in this limit, $\Gamma_1>0$ as $\eta=\Omega/\omega\to\infty$, instead means that the symmetry-broken states have a finite lifetime $\tau\sim 1/\Gamma_1$, after which the system relaxes to the unique, symmetry-restored steady state. The second decay rate $\Gamma_2$ is the decay rate of the next slowest mode. The ratio $\Gamma_2/\Gamma_1\geq 1$ quantifies the separation between the slowest relaxation timescale and all the others. When $\Gamma_2\gg\Gamma_1$, the lifetime $1/\Gamma_1$ is much longer than all other relaxation timescales, and the state is called metastable~\cite{macieszczak2016}. In finite-component light--matter systems, such a metastable superradiant state was recently identified in the dissipative isotropic quantum Rabi model~\cite{xiao2026metastability}, showing that dynamical, not just steady-state, signatures are needed to characterize dissipative criticality in these systems. 

In this work, we investigate the anisotropic open Rabi model in the simultaneous presence of cavity photon loss, atomic relaxation, and atomic dephasing. Combining semiclassical stability analysis~\cite{xiao2026metastability} with exact Liouvillian diagonalization, we determine the steady-state phase diagram, the Liouvillian spectrum, and the phase-space structure of the steady state via Wigner distribution function~\cite{xiao2026metastability,lyu2024multicritical, xu2019emergent}.

Our central result is that atomic relaxation and atomic dephasing act as competing, 
qualitatively distinct resources for superradiant order, distinguishable both in the 
steady-state phase space and in the Liouvillian spectrum. Deep in the superradiant phase, 
cavity photon loss supports a genuine dissipative phase transition. The Liouvillian gap 
closes algebraically with the atomic-to-cavity frequency ratio, while the Wigner function exhibits two well-separated peaks in phase space. Adding atomic dephasing to cavity photon loss changes the nature of this 
transition. The mean-field superradiant fixed points vanish entirely, 
the Wigner-function peaks progressively lose weight to a single central peak, and the Liouvillian gap shows finite value with the ratio $\Gamma_2/\Gamma_1 \approx 1$, signaling convergence 
towards a unique normal-phase steady state. In contrast, adding 
atomic relaxation to cavity photon loss arrests this closing: the Liouvillian gap saturates to a finite 
value, while the Wigner function develops a persistent, interference-textured channel linking the two peaks, signaling that population coherently and repeatedly transits between them. The superradiant phase observed here at finite frequency ratios is therefore not a genuine dissipative phase transition but a long-lived metastable state.

These results show that dissipative criticality in finite-component light--matter systems 
depends not only on the strength of the environmental coupling but on its microscopic origin. 
Whether a dissipation channel recycles atomic population, as atomic relaxation does, or 
merely erases atomic coherence, as atomic dephasing does, determines whether it sustains or 
shortens long-lived superradiant dynamics.

\section{Model}
\label{sec:model}
We study a single two-level atom coupled to a single-mode optical cavity, a system that serves as the minimal arena for exploring dissipative phase transitions in quantum light--matter systems. The coherent part of the dynamics is governed by the anisotropic quantum Rabi 
Hamiltonian~\cite{xie2014anisotropic, lyu2024multicritical},
\begin{equation}
    \hat{H} = \frac{\Omega}{2}\hat{\sigma}_{z} 
    + \omega\hat{a}^{\dagger}\hat{a}
    - \frac{\lambda_{x}}{2}
      (\hat{a}+\hat{a}^{\dagger})\hat{\sigma}_{x}
    - \frac{i\lambda_{y}}{2}
      (\hat{a}-\hat{a}^{\dagger})\hat{\sigma}_{y},
    \label{eq:hamiltonian}
\end{equation}
\begin{figure}[t]
\includegraphics[width=\linewidth]{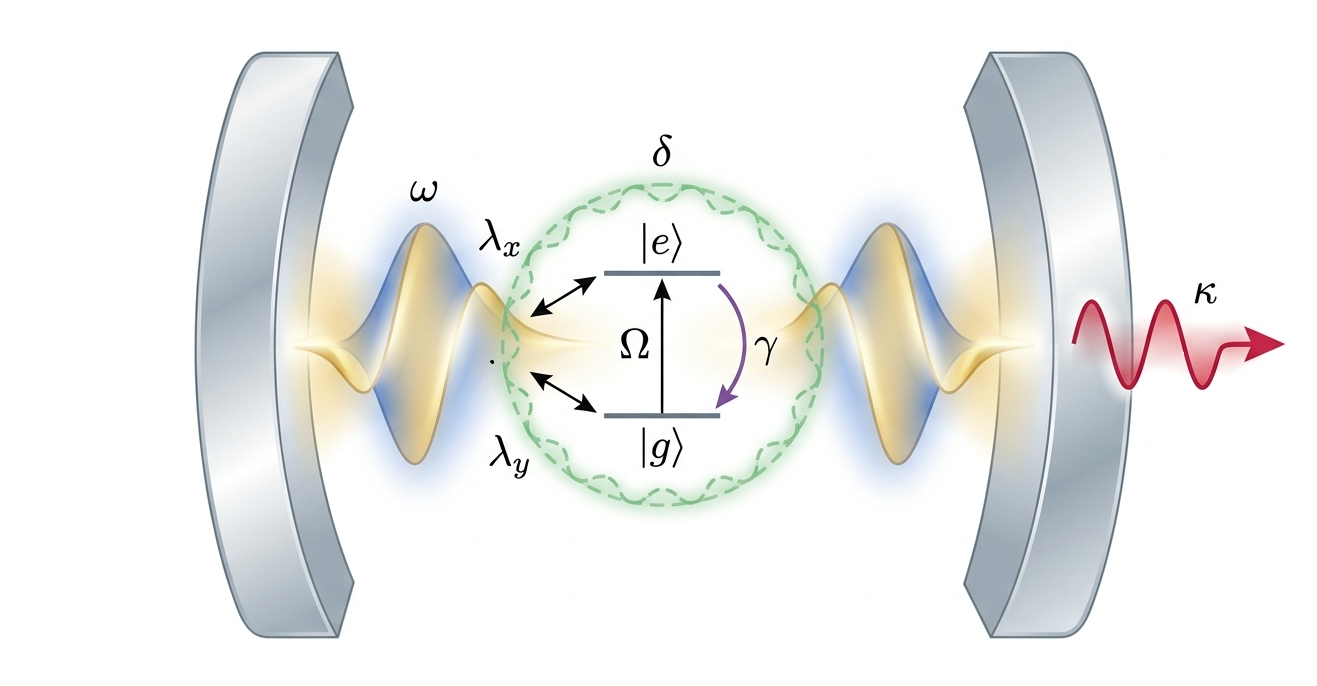}
\caption{Schematic diagram of the anisotropic Rabi system in an open cavity QED setting. A two-level atom with ground state $|g\rangle$ and excited state $|e\rangle$ interacts with a single cavity mode of frequency $\omega$. Photon leakage through the cavity mirrors occurs at rate $\kappa$, atomic relaxation at rate $\gamma$, and pure dephasing at rate $\delta$.}
\label{fig:schematic}
\end{figure}
where $\hat{\sigma}_{x,y,z}$ are the Pauli operators of the two-level atom with transition frequency $\Omega$, $\hat{a}$ ($\hat{a}^{\dagger}$) are the bosonic annihilation (creation) operators of the cavity mode with frequency $\omega$, and $\lambda_{x}$, $\lambda_{y}$ are the coupling strengths along the position- and momentum-like quadratures of the field, $\hat{x} \propto (\hat{a}+\hat{a}^{\dagger})$ and $\hat{p} \propto i(\hat{a}^{\dagger}-\hat{a})$, respectively.

This form of the Hamiltonian interpolates between the Jaynes–Cummings model, the standard isotropic Rabi model, and strongly anisotropic regimes depending on the relative values of $\lambda_x$ and $\lambda_y$. In the limit $\lambda_{y} \to 0$ it recovers the isotropic quantum Rabi model~\cite{bishop1996application, bishop2001time, levine2004entanglement, hwang2010variational, yu2012analytical, bakemeier2012quantum, ashhab2013superradiance, hwang2015quantum, wei2018fidelity, sun2020out, zhang2021quantum}, in which rotating and counter-rotating processes enter symmetrically. Setting $\lambda_{x} = \lambda_{y} \equiv \lambda$, it reduces to the Jaynes--Cummings model~\cite{larson2007dynamics, hwang2016quantum}, where only energy-conserving (rotating-wave) processes survive. For general $\lambda_{x} \neq \lambda_{y}$, the two types of interaction are present with unequal weight, defining the anisotropic regime~\cite{xie2014anisotropic, liu2017universal, wang2018quantum, chen2021multiple} which is the focus of this work. 

We model the environment through three physically distinct dissipation channels, each of which acts on a different degree of freedom and plays a qualitatively different role in the nonequilibrium dynamics. The open-system evolution is governed by the Lindblad master equation~\cite{kirton2017suppressing}
\begin{equation}
    \frac{d\hat{\rho}}{dt} = -i[\hat{H},\hat{\rho}]
    + \kappa\,\mathcal{D}[\hat{a}]\hat{\rho}
    + \gamma\,\mathcal{D}[\hat{\sigma}_{-}]\hat{\rho}
    + \delta\,\mathcal{D}[\hat{\sigma}_{z}]\hat{\rho},
    \label{eq:master}
\end{equation}
which can be written compactly in terms of the Liouvillian superoperator $\mathcal{L}$ as
\begin{equation}
\frac{d\hat{\rho}}{dt} = \mathcal{L}\hat{\rho}.
\label{eq:liouvillian}
\end{equation}
Here the dissipator associated with a jump operator $\hat{L}$ is defined as
\begin{equation}
\mathcal{D}[\hat{L}]\hat{\rho}
= 2\hat{L}\hat{\rho}\hat{L}^{\dagger}
- \hat{L}^{\dagger}\hat{L}\hat{\rho}
- \hat{\rho}\hat{L}^{\dagger}\hat{L},
\label{eq:dissipator}
\end{equation}
and $\hat{\sigma}_{\pm} = (\hat{\sigma}_{x} \pm i\hat{\sigma}_{y})/2$ are the atomic raising and lowering operators.
The schematic diagram of this model is shown in Fig.~\ref{fig:schematic}. The three channels in Eq.~\eqref{eq:master} are physically distinct in the way they couple to the superradiant order. Cavity photon loss at rate $\kappa$ with jump operator $\hat{a}$ describes photon leakage through the cavity mirrors and damps the cavity field amplitude. Atomic relaxation at rate $\gamma$, with jump operator $\hat{\sigma}_{-}$, incoherently relaxes the atom to its ground state by continuously recycling atomic population back into the ground state. Atomic dephasing at rate $\delta$, with jump operator $\hat{\sigma}_{z}$, randomizes the phase between the ground state and the excited state of the atom without any population transfer. Since the superradiant order parameter is sustained precisely by this atomic coherence, atomic dephasing acts as a direct adversary of superradiance in a way that atomic relaxation does not.

Using $\hat{\sigma}_{z}^{2} = \mathbbm{1}$, the dephasing 
dissipator simplifies to
\begin{equation}
    \mathcal{D}[\hat{\sigma}_{z}]\hat{\rho}
    = 2\hat{\sigma}_{z}\hat{\rho}\hat{\sigma}_{z} 
      - 2\hat{\rho},
    \label{eq:dephasing_simplified}
\end{equation}
which makes transparent that dephasing destroys the off-diagonal elements of the atomic density matrix while leaving the populations unchanged. This distinction has direct consequences for the phase structure. In Sec.~\ref{sec:meanfield}, we show that combining cavity photon loss with atomic dephasing ($\kappa + \delta$) destabilizes the superradiant phase entirely, whereas combining cavity photon loss with atomic relaxation ($\kappa + \gamma$) preserves the qualitative topology of the phase diagram. The physical origin of this asymmetry lies precisely in the different ways in which these two atomic channels interact with the coherence that underpins superradiant order.

\section{Mean-field theory and stability analysis}
\label{sec:meanfield}

\subsection{Equations of motion}
We derive the semiclassical equations of motion by applying the mean-field decoupling $\langle \hat{A}\hat{B} \rangle \approx \langle\hat{A}\rangle\langle\hat{B}\rangle$ to the Heisenberg equations of motion generated by the master equation given in Eq.~\eqref{eq:master}. This approximation becomes exact in the thermodynamic limit $\eta \equiv \Omega/\omega \to \infty$, which we adopt throughout this section. The resulting equations of motion for $\langle\hat{a}\rangle$, $\langle\hat{\sigma}_{+}\rangle$, and $\langle\hat{\sigma}_{z}\rangle$ are

\begin{widetext}
\begin{align}
\langle \dot{\hat{a}} \rangle &=
(-i\omega - \kappa)\langle \hat{a} \rangle
+ \frac{i}{2}(\lambda_x - \lambda_y)
  \langle \hat{\sigma}_+ \rangle
+ \frac{i}{2}(\lambda_x + \lambda_y)
  \langle \hat{\sigma}_- \rangle,
\label{eq:eom_a}
\\
\langle \dot{\hat{\sigma}}_+ \rangle &=
\frac{i}{2}(\lambda_ x - \lambda_y)\langle\hat{a}\rangle\langle\hat{\sigma_z}\rangle
+\frac{i}{2}(\lambda_x + \lambda_y)
  \langle\hat{a}^\dagger\rangle \langle\hat{\sigma_z}\rangle
+ (i\Omega - \gamma - 4\delta)
  \langle \hat{\sigma}_+ \rangle,
\label{eq:eom1_sp}
\\
\langle \dot{\hat{\sigma}}_z \rangle &=
i\lambda_x 
(\langle\hat{a}\rangle + \langle\hat{a}^\dagger\rangle)
(\langle\hat{\sigma}_+\rangle - 
 \langle\hat{\sigma}_-\rangle)
+i\lambda_y
(\langle\hat{a}\rangle - \langle\hat{a}^\dagger\rangle)
(\langle\hat{\sigma}_+\rangle + 
 \langle\hat{\sigma}_-\rangle)
-2\gamma(1 + \langle\hat{\sigma}_z\rangle).
\label{eq:eom1_sz}
\end{align}
\end{widetext}
Two features of these equations are worth noting. First, atomic relaxation ($\gamma$) and pure dephasing ($\delta$) both enter the coherence decay in Eq.~\eqref{eq:eom1_sp}, but through different mechanisms: $\gamma$ damps the coherence while simultaneously driving population relaxation in Eq.~\eqref{eq:eom1_sz}, whereas $\delta$ damps only the coherence with no accompanying population dynamics. This asymmetry is the microscopic origin of the qualitatively different roles that these two channels play in the phase diagram. Second, the combination $\gamma + 4\delta$ that governs coherence decay in Eq.~\eqref{eq:eom1_sp} follows directly from the simplification $\hat{\sigma}_z^2 = \mathbbm{1}$ applied to the dephasing dissipator in Eq.~\eqref{eq:dephasing_simplified}.

It is convenient to introduce dimensionless variables, following the rescaling scheme of 
Ref.~\cite{lyu2024multicritical}. We rescale the cavity amplitude and the coupling strengths as
\begin{equation}
    \langle\hat{a}\rangle = \alpha\sqrt{\eta},
    \quad
    \tilde{\lambda}_{x,y} = 
    \frac{\lambda_{x,y}}{\sqrt{\Omega\omega}},
    \quad
    s_\pm = \langle\hat{\sigma}_\pm\rangle,
    \quad
    s_z = \langle\hat{\sigma}_z\rangle,
    \label{eq:rescaling}
\end{equation}
and define the three dimensionless dissipation rates
\begin{equation}
    \tilde{k} = \frac{\kappa}{\omega},
    \qquad
    \Gamma  = \frac{\gamma + 4\delta}{\Omega},
    \qquad
    \Gamma_0 = \frac{\gamma}{\Omega},
    \label{eq:dimless_rates}
\end{equation}
where $\Gamma$ governs coherence decay and $\Gamma_0$ governs population relaxation, so 
that $\Gamma=\Gamma_0$ when $\delta=0$, and $\Gamma>\Gamma_0$ whenever dephasing is 
present. The mean field equations become:

\begin{widetext}
\begin{align}
\dot{\alpha} &=
-i\omega \left[ \left( 1- \frac{i\kappa}{\omega} \right) \alpha - \frac{1}{2}\left( \tilde{\lambda}_x (s_+ +s_+^*) - \tilde{\lambda}_y (s_+ -s_+^*)\right) \right],
\label{eq:eom_a}
\\
\dot{s}_+ &= \frac{i\Omega}{2}\left[ \tilde{\lambda}_x (\alpha + \alpha^*) - \tilde{\lambda}_y (\alpha - \alpha^*) \right] s_z + i\Omega \left( 1+i\Gamma \right)s_+,
\label{eq:eom_sp}
\\
\dot{s}_z &= i\Omega \left[ \tilde{\lambda}_x (\alpha + \alpha^*) (s_+ -s_+^*)+ \tilde{\lambda}_y (\alpha - \alpha^*) (s_+ + s_+^*) + 2i\Gamma_0 \left( 1+s_z \right)\right].
\label{eq:eom_sz}
\end{align}
\end{widetext}

Decomposing the complex amplitudes into real and 
imaginary parts,
\begin{equation}
    \alpha = x + iy, \qquad 
    s_+ = u + iv, \qquad 
    s_z = z,
    \label{eq:decomp}
\end{equation}
the mean-field equations reduce to the five-dimensional 
real dynamical system
\begin{align}
    \dot{x} &= \omega(y - \tilde{k}x 
               + \tilde{\lambda}_y v),
    \label{eq:x}\\
    \dot{y} &= \omega(-x - \tilde{k}y 
               + \tilde{\lambda}_x u),  
    \label{eq:y}\\
    \dot{u} &= -\Omega(\Gamma u + v 
               - \tilde{\lambda}_y yz),
    \label{eq:u}\\
    \dot{v} &= \Omega(u - \Gamma v 
               + \tilde{\lambda}_x xz),
    \label{eq:v}\\
    \dot{z} &= -4\Omega(\tilde{\lambda}_x xv 
               + \tilde{\lambda}_y yu) 
               - 2\Omega\Gamma_0(1+z).
    \label{eq:z}
\end{align}
The steady states of Eqs.~\eqref{eq:x}--\eqref{eq:z} determine the phase diagram. The system always admits the trivial solution $(x,y,u,v,z) = (0,0,0,0,-1)$, corresponding to the NP with no cavity field and the atom in its ground state. Nontrivial solutions with $\alpha \neq 0$ correspond to the SP and, when they exist, appear in $\mathbb{Z}_2$-symmetry-related pairs $(x_*,y_*,u_*,v_*,z_*)$ and $(-x_*,-y_*,-u_*,-v_*,z_*)$, reflecting the spontaneous breaking of the parity symmetry $\hat{P} = \exp[i\pi(\hat{a}^\dagger\hat{a} + (\hat{\sigma}_z +1)/2)]$ of the Hamiltonian.

\subsection{Normal-phase stability}
\label{subsec:NP}
The stability of the NP is determined by linearizing Eqs.~\eqref{eq:x}--\eqref{eq:z} around the fixed point $(0,0,0,0,-1)$. Since $\dot{z}$ evaluated at this point decouples from the cavity and spin-coherence variables — reducing to $\dot{z} = -2\Omega\Gamma_0(1+z)$, which is unconditionally stable — the NP stability is governed entirely by the $4\times 4$ linearized subsystem for $(x,y,u,v)$. Setting the corresponding coefficient matrix to zero,
\begin{equation}
    \mathbf{M}
    \begin{pmatrix} x \\ y \\ u \\ v \end{pmatrix}
    = \mathbf{0},
    \qquad
    \mathbf{M} =
    \begin{pmatrix}
        -\tilde{k} & 1 & 0 & \tilde{\lambda}_y \\
        -1 & -\tilde{k} & \tilde{\lambda}_x & 0 \\
        0 & -\tilde{\lambda}_y & -\Gamma & -1\\
        -\tilde{\lambda}_x & 0 & 1 & -\Gamma 
        
    \end{pmatrix},
    \label{eq:M}
\end{equation}
a nontrivial solution requires $\det\mathbf{M} = 0$.  
Evaluating the determinant gives the NP phase boundary
\begin{equation}
    \det\mathbf{M} = 
    \tilde{k}^2 + \Gamma^2(1+\tilde{k}^2)
    + 2\Gamma\tilde{k}
      \tilde{\lambda}_x\tilde{\lambda}_y
    + (1-\tilde{\lambda}_x^2)(1-\tilde{\lambda}_y^2) 
    = 0,
    \label{eq:NP_boundary}
\end{equation}
which is an exact analytical result for the instability boundary in the $(\tilde{\lambda}_x,\tilde{\lambda}_y)$ plane, parametrized by $\tilde{k}$, $\Gamma$, and $\Gamma_0$. The NP is stable (unstable) for $\det\mathbf{M} > 0$ ($< 0$).

Eq.~\eqref{eq:NP_boundary} has an important limiting structure. In the absence of atomic dissipation ($\Gamma \to 0$), it reduces to
\begin{equation}
    \tilde{k}^2 + 
    (1-\tilde{\lambda}_x^2)(1-\tilde{\lambda}_y^2) = 0,
    \label{eq:NP_cavity_only}
\end{equation}
recovering the known result for cavity photon loss alone ~\cite{hwang2018dissipative}. For finite $\Gamma$, the cross-term $2\Gamma\tilde{k}\tilde{\lambda}_x\tilde{\lambda}_y$ 
breaks the symmetry between $\tilde{\lambda}_x$ and $\tilde{\lambda}_y$, deforming the phase 
boundary in a way that depends explicitly on the product of both coupling strengths. Setting 
$\tilde{\lambda}_y=0$ (or, by the $x\leftrightarrow y$ symmetry of 
Eq.~\eqref{eq:NP_boundary}, $\tilde{\lambda}_x=0$), Eq.~\eqref{eq:NP_boundary} yields the 
critical coupling
\begin{equation}
    \tilde{\lambda}_{x,y}^{\mathrm{c}} = 
    \sqrt{1+\frac{\kappa^2}{\omega^2}}
    \sqrt{1+\frac{(\gamma+4\delta)^2}{\Omega^2}},
    \label{eq:lambda_crit}
\end{equation}
showing that both cavity photon loss and atomic dissipations shift the critical point to larger coupling, consistent with their role as competing channels that must be overcome to sustain superradiant order.

\subsection{Superradiant-phase stability}
\label{subsec:SP}
Beyond the NP boundary, the system can support nontrivial fixed points with $\alpha \neq 0$. When only cavity photon loss is present ($\Gamma_0 = 0$, $\Gamma = 0$), the Bloch-sphere constraint $4(u^2+v^2)+z^2 = 1$ holds and the SP fixed points can be found analytically~\cite{kirton2019introduction}. The addition of atomic dephasing breaks this constraint: atomic relaxation drives incoherent population relaxation that mixes the spin's pure state, and atomic dephasing further randomizes its phase, so that $(x,y,u,v,z)$ must be treated as five fully coupled variables. The SP fixed points are therefore determined by solving the nonlinear system Eqs.~\eqref{eq:x}--\eqref{eq:z} set to zero numerically at each point in parameter space, using the NP solution as a seed and tracking solution branches across the phase boundary.

The stability of each fixed point is assessed via its Jacobian,
\begin{equation}
\small
\setlength{\arraycolsep}{2pt}
\renewcommand{\arraystretch}{1.12}
\mathbf{J}\big|_{*}=
\begin{pmatrix}
-\omega\tilde{k} & \omega & 0 & \omega\tilde{\lambda}_y & 0 \\
-\omega & -\omega\tilde{k} & \omega\tilde{\lambda}_x & 0 & 0 \\
0 & \Omega\tilde{\lambda}_y z_*  & -\Omega\Gamma & -\Omega & \Omega\tilde{\lambda}_y y_* \\
\Omega\tilde{\lambda}_x z_* & 0 & \Omega & -\Omega\Gamma & \Omega\tilde{\lambda}_x x_* \\
-4\Omega\tilde{\lambda}_x v_* & -4\Omega\tilde{\lambda}_y u_* &
-4\Omega\tilde{\lambda}_y y_* & -4\Omega\tilde{\lambda}_x x_* &
-2\Omega\Gamma_0
\end{pmatrix}.
\label{eq:jac_sp}
\end{equation}
A fixed point is stable if and only if all eigenvalues $\{\mu_i\}$ of $\mathbf{J}|_*$ satisfy $\mathrm{Re}(\mu_i) < 0$; the SP phase boundary is the locus at which $\max_i\,\mathrm{Re}(\mu_i) = 0$. One can verify that evaluating $\mathbf{J}|_*$ at the NP fixed point $(0,0,0,0,-1)$ recovers exactly the matrix $\mathbf{M}$ in the upper $4\times 4$ block, with the fifth mode decoupling as $\mu_5 = -2\Omega\Gamma_0 < 0$. Equation~\eqref{eq:NP_boundary} is thus a special case of the general eigenvalue condition, confirming the internal consistency of the two approaches.
\begin{figure}[t]
\centering
\includegraphics[width=\columnwidth]{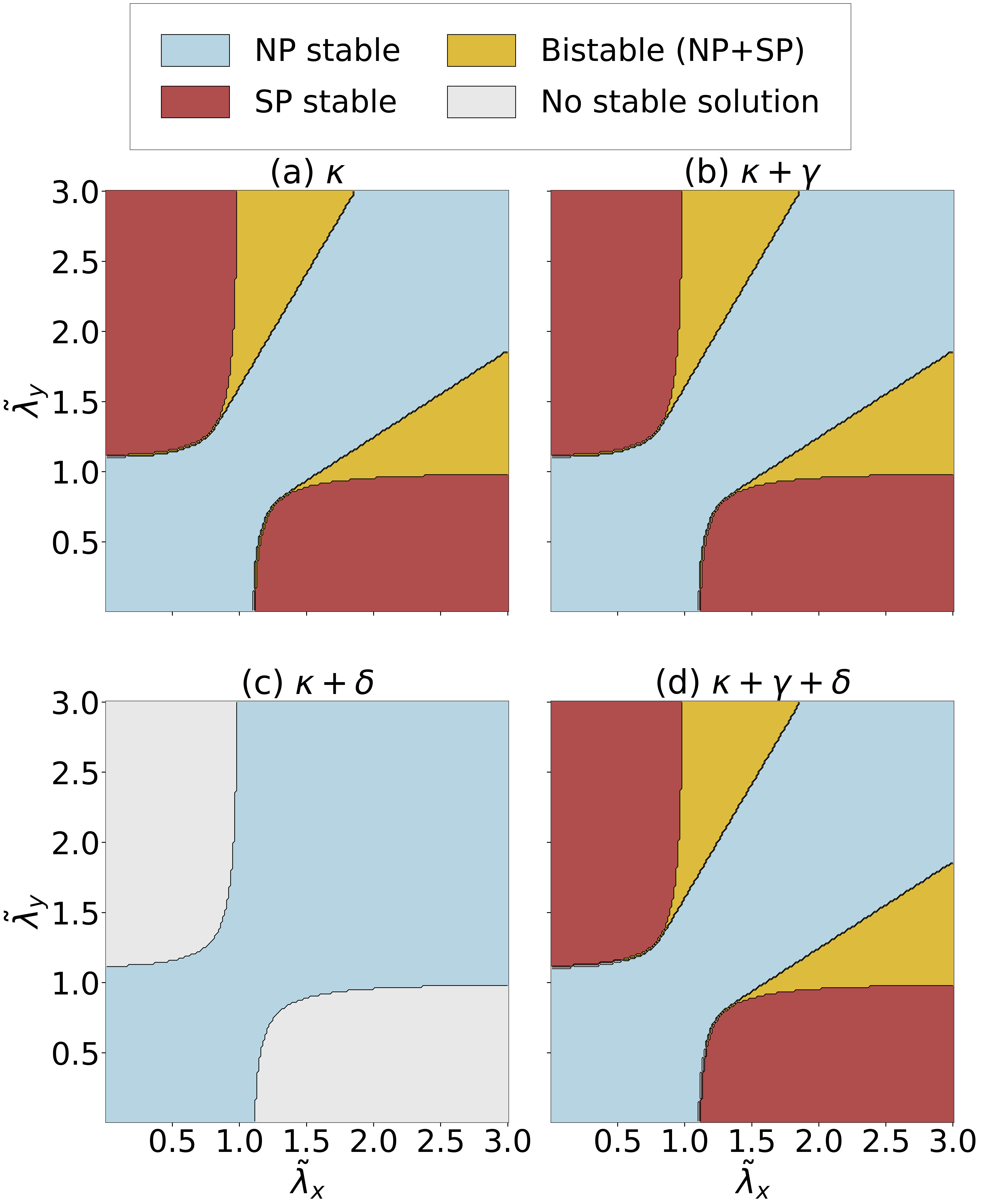}
\caption{Phase diagrams in the $(\tilde{\lambda}_{x}, \tilde{\lambda}_{y}) $ plane for different combinations of dissipation. Blue regions: stable normal phase (NP); green: stable superradiant phase (SP); yellow: bistable (NP+SP); white: no stable fixed point. Solid lines show analytic NP boundaries, dashed lines the numerically determined SP boundaries. The parameters taken here are $\omega = 1$, $\Omega = 100$, $\kappa = 0.5$, $\gamma = 0.2, \delta = 0.01$.}
\label{fig:analytic_boundary}
\end{figure}

\begin{figure*}[t]
    \includegraphics[width=0.95\textwidth]{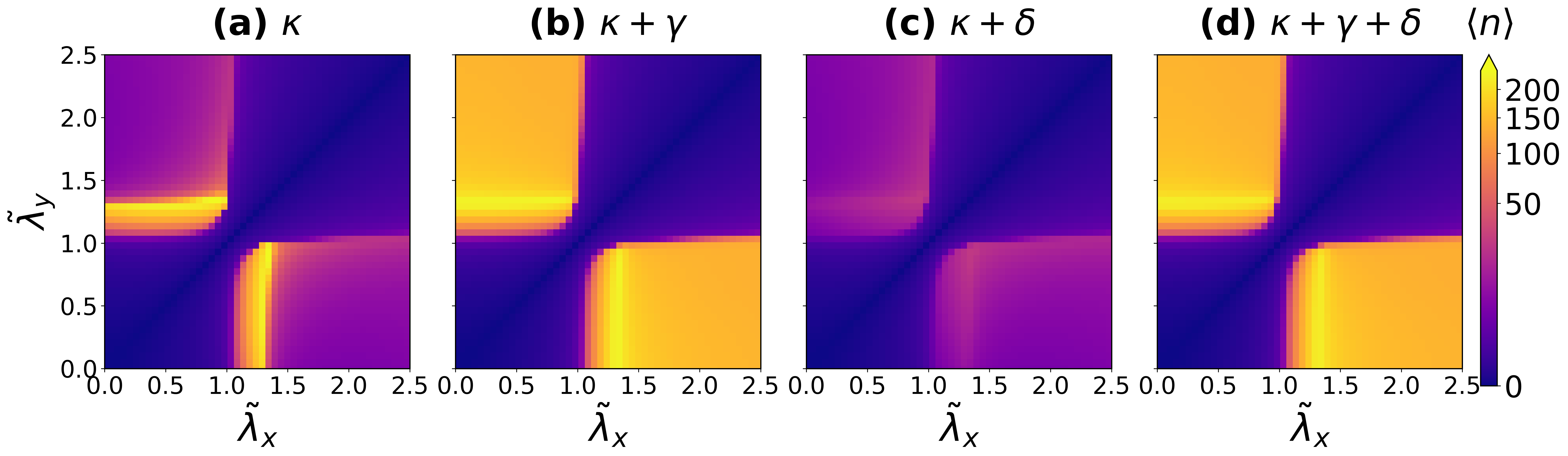}
    \caption{Steady-state photon number \(\langle a^\dagger a \rangle\) obtained from exact Liouvillian diagonalization, plotted over the \((\tilde{\lambda}_x, \tilde{\lambda}_y)\) plane for different combinations of cavity decay, atomic relaxation, and dephasing. The parameters taken here are $\omega = 1$, $\Omega = 1000$, $\kappa = 0.5$, $\gamma = 0.2, \delta = 0.01$. The bosonic cutoff is: $N_{\text{cutoff}}= 300$.}
    \label{fig:steady_state_photon}
\end{figure*}
\subsection{Phase diagram}
\label{subsec:phasediagram}
The full phase diagram in the $(\tilde{\lambda}_x,\tilde{\lambda}_y)$ plane is assembled by 
combining the analytical NP boundary~\eqref{eq:NP_boundary} with the numerical SP stability 
analysis above, shown in Fig.~\ref{fig:analytic_boundary} for four representative dissipation 
configurations.

For cavity photon loss alone [$\kappa=0.5$, $\gamma=\delta=0$; Fig.~\ref{fig:analytic_boundary}(a)], 
the phase diagram reproduces the structure established for the anisotropic Rabi 
model in Refs.~\cite{soriente2018dissipation, lyu2024multicritical}: a second-order NP--SP boundary connects two 
tricritical points, beyond which first-order boundaries enclose a bistable region where NP 
and SP are simultaneously stable. Adding atomic relaxation [$\gamma=0.2$; 
Fig.~\ref{fig:analytic_boundary}(b)] and, further, atomic dephasing 
[$\delta=0.01$; Fig.~\ref{fig:analytic_boundary}(d)] each preserve this topology, merely 
shifting the boundaries. Here population recycling by 
atomic relaxation maintains the atomic coherence needed to sustain the SP.

The combination $\kappa+\delta$ with $\gamma=0$ [Fig.~\ref{fig:analytic_boundary}(c)] tells a 
different story: the stable SP is completely absent from the phase diagram. Atomic dephasing, 
by destroying atomic coherence without recycling population, removes the resource the 
superradiant phase requires to exist at the mean-field level. 
\subsection{Quantum Order Parameters and Comparison with Mean-Field Theory}
\label{subsec:order_parameter}
The mean-field analysis presented in Sec.~\ref{sec:meanfield} A identifies regions in the $(\tilde{\lambda}_x,\tilde{\lambda}_y)$ parameter space where the NP, SP, and their bistable coexistence BP emerge as locally stable solutions of the semiclassical equations of motion. To verify whether these phases persist in the full quantum description of the dissipative system, we compute steady-state expectation values of relevant photonic and spin observables from the exact Liouvillian steady state $\rho_{\mathrm{ss}}$.

Within the Lindblad formalism, the density matrix $\rho(t)$ of the coupled atom--cavity system evolves according to the master equation
\begin{equation}
\frac{d\rho(t)}{dt}
=
\mathcal{L}\rho(t),
\label{eq:lindblad_master}
\end{equation}
where $\mathcal{L}$ denotes the Liouvillian superoperator introduced in Eq.~(\ref{eq:liouvillian}). The steady state $\rho_{\mathrm{ss}}$ is defined as the time-independent solution of Eq.~(\ref{eq:lindblad_master}), satisfying
\begin{equation}
\mathcal{L}\rho_{\mathrm{ss}}=0.
\label{eq:steady_state_condition}
\end{equation}
 Physically, it describes the stationary state of the dissipative anisotropic Rabi model and incorporates both the coherent correlations generated by the system Hamiltonian and the fluctuations induced by the dissipative channels. Consequently, all steady-state observables can be evaluated directly from $\rho_{\mathrm{ss}}$. 
In particular, the expectation value of an operator $\hat{O}$ is given by
\begin{equation}
\langle \hat{O} \rangle
=
\mathrm{Tr}
\!\left(
\rho_{\mathrm{ss}}\hat{O}
\right).
\label{eq:observable_expectation}
\end{equation}
Using this relation, we calculate the steady-state photon number
\begin{equation}
n_{\mathrm{ph}}
=
\langle a^{\dagger}a\rangle,
\end{equation}
and in Fig.~\ref{fig:steady_state_photon} we plot it for four different cases ($(a)\ \kappa\neq 0,\ \gamma,\delta=0,\ (b)\ \kappa,\gamma\neq 0,\ \delta=0,\ (c)\ \kappa,\delta\neq 0,\ \gamma=0,\ (d)\ \kappa,\gamma,\delta\neq 0$) as in the previous section. This quantity serves as a quantum order parameter that allows for a direct comparison between the exact steady-state properties of the open system and the semiclassical phase diagram obtained from the mean-field analysis (Sec.~\ref{sec:meanfield}).

 The calculations confirm the existence and location of the normal and superradiant phases, while also revealing where the exact treatment departs from the idealized semiclassical picture. Sharp mean-field phase boundaries are replaced by smooth crossovers in $n_{\mathrm{ph}}$, and, near the bistable region, quantum tunneling between the NP- and SP-like configurations mixes their steady-state weight rather than allowing them to coexist as strictly independent fixed points.

\section{Phase-space structure of the steady state}
\label{sec:wigner}
The stability analysis of Sec.~\ref{sec:meanfield} provides a semiclassical skeleton of the model by identifying the NP, the SP, the BP and their regions of local stability. However, it does not reveal how quantum fluctuations affect the surrounding phase space or whether the quantum steady state is localized near a single classical attractor or spreads over a manifold connecting several of them. To address this, we examine the Wigner quasiprobability distribution of the cavity field.

\begin{figure*}[t]
    \includegraphics[width=0.88\textwidth]{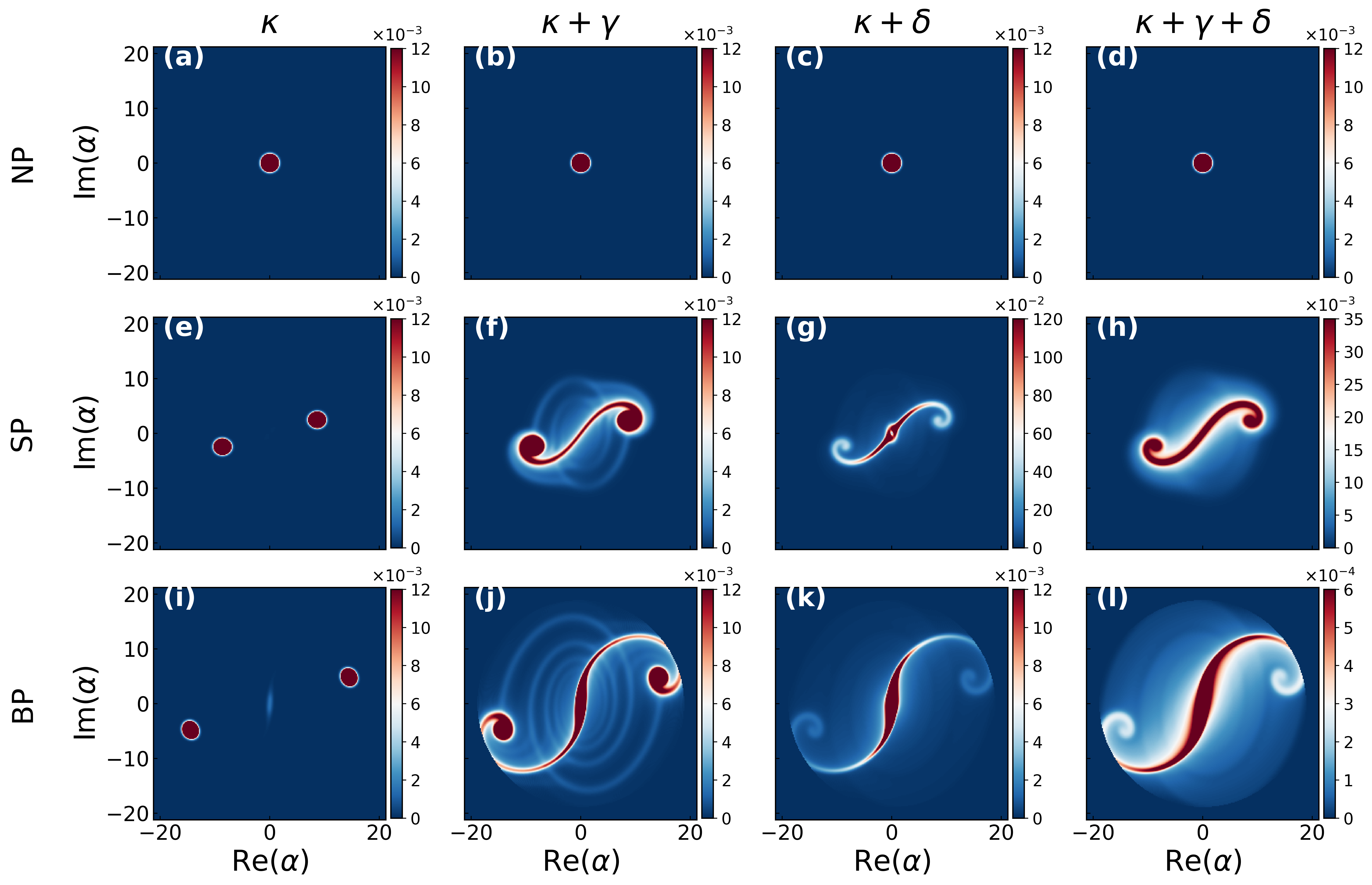}
    \caption{Wigner functions of the cavity field in the normal phase (NP), superradiant phase (SP), and bistable phase (BP), shown in the complex amplitude plane \(\alpha=\operatorname{Re}(\alpha)+i\,\operatorname{Im}(\alpha)\). The dimensionless light--matter coupling parameters are \((\tilde{\lambda}_x,\tilde{\lambda}_y)=(0.5,\,0.5)\) for the NP, \((\tilde{\lambda}_x,\tilde{\lambda}_y)=(1.5,\,0.5)\) for the SP, and \((\tilde{\lambda}_x,\tilde{\lambda}_y)=(2.3,\,1.1)\) for the BP. The remaining parameters are \(\kappa/\omega=0.5\), \(\gamma/ \omega=0.5\), and \(\delta / \omega=0.5\). The bosonic cutoff is: $N_{\text{cutoff}}= 400$.}
    \label{fig:wigner_combined}
\end{figure*}

Starting from the steady-state density matrix $\rho_{\mathrm{ss}}$, we trace out the spin and construct the reduced cavity state
\begin{equation}
  \rho_{\mathrm{cav}} = \mathrm{Tr}_{\mathrm{spin}}\!\left[\rho_{\mathrm{ss}}\right].
  \label{eq:rhocav}
\end{equation}
The corresponding Wigner function is calculated as
\begin{equation}
W(\alpha) = \frac{2}{\pi}\text{Tr}\left[\hat{\rho}_{\rm cav}\hat{D}(\alpha)(-1)^{\hat{a}^\dagger\hat{a}}\hat{D}^\dagger(\alpha)\right],
\label{eq:Wigner}
\end{equation}
with displacement operator
\begin{equation}
  D(\alpha) = \exp\!\left(\alpha a^{\dagger} - \alpha^{*} a\right),
\end{equation}
and coordinate of the phase-space $\alpha = (x+ip)/\sqrt{2}$. The results are shown in Fig.~\ref{fig:wigner_combined} for representative points ($\tilde{\lambda}_x, \tilde{\lambda}_y $) in the NP, SP, and BP regions and for the four dissipation configurations. 

\subsection{Normal and superradiant phases}
The top row of Fig.~\ref{fig:wigner_combined} shows $W(\alpha)$ in the NP for all dissipation combinations. In every case the distribution is a single nearly Gaussian peak centered at the origin, reflecting weak fluctuations around the vacuum (see Fig.~\ref{fig:wigner_combined}(a - d)).

The middle row (see Fig.~\ref{fig:wigner_combined}(e - h)) displays the Wigner functions in SP. For cavity photon loss only [Fig.~\ref{fig:wigner_combined}(e)] the distribution consists of two sharply localized peaks at $\pm\alpha_{\mathrm{SP}}$ with negligible weight in between, mirroring the pair of symmetry-broken semiclassical attractors. When atomic relaxation is added to cavity photon loss[Fig.~\ref{fig:wigner_combined}(f)] a bright S-shaped bridge develops between the two peaks and faint spiral fringes appear, showing that atomic relaxation induces rare excursions along the superradiant loop in phase space. Adding atomic dephasing with cavity photon loss [Fig.~\ref{fig:wigner_combined}(g)] collapses the two symmetry-broken peaks toward a single, sharply peaked feature near the origin, indicating that dephasing does not 
merely wash out coherence locally but erodes the superradiant order parameter itself. And with all three channels present [Fig.~\ref{fig:wigner_combined}(h)] the probability weight is redistributed along a continuous S-shaped band connecting the two peaks. The steady state is then strongly non-Gaussian and cannot be described by a single coherent state.

\subsection{Bistable region}
The bottom row of Fig.~(\ref{fig:wigner_combined}) shows Wigner functions at a representative 
point in the bistable region, where mean-field theory predicts coexistence of NP-like and 
SP-like fixed points. For cavity decay alone [Fig.~\ref{fig:wigner_combined}(i)], three 
distinct features are visible: a faint peak near the origin and two sharp, isolated peaks at 
$\pm\alpha_{\mathrm{SP}}$, with no connecting bridge. The steady state is well approximated 
by a statistical mixture of NP and SP configurations with only weak residual coupling between 
them.

Adding atomic relaxation [Fig.~\ref{fig:wigner_combined}(j)] connects the three peaks via 
a pronounced, fringe-rich S-shaped ridge with sizeable intermediate weight, indicating that 
the population dynamically and repeatedly transits the entire hysteresis loop rather than 
remaining localized near the three semiclassical fixed points. Superimposing atomic  dephasing to cavity photon loss
[Fig.~\ref{fig:wigner_combined}(k)] weakens the surrounding 
interference fringes, and shifts weight toward the central, NP-like region which is consistent with 
dephasing suppressing coherent superradiant order. With both atomic channels and cavity photon loss present [Fig.~\ref{fig:wigner_combined}(l)], the quasiprobability spreads along the loop, with no sharp maxima at either the former NP or SP 
locations.

In summary, atomic dissipation qualitatively reshapes the phase-space structure of the steady state: in the SP region, cavity loss alone yields two well-separated peaks, while atomic relaxation and dephasing progressively populate the connecting superradiant loop; in the bistable region, cavity decay alone supports a trimodal NP+SP mixture, while atomic relaxation and dephasing together produce a strongly delocalized, interference-textured distribution.

\begin{figure*}[t]          
    \centering
    \includegraphics[width=0.80\textwidth]{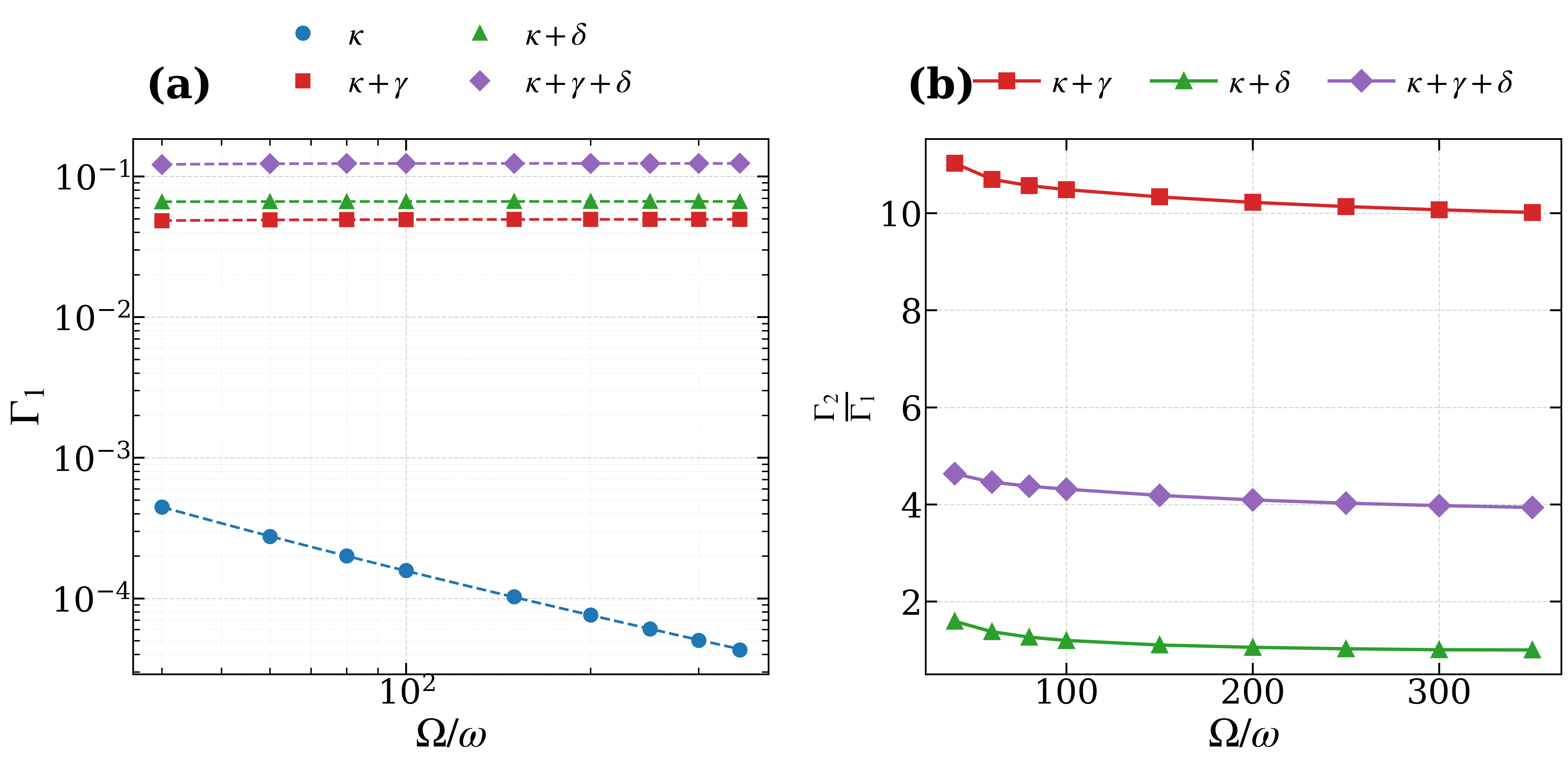}
    \caption{(a) Slowest decay rate $\Gamma_1$ and (b) ratio $\Gamma_2/\Gamma_1$ versus $\eta=\Omega/\omega$ in the superradiant phase, $(\tilde\lambda_x,\tilde\lambda_y)=(1.5,0.5)$, for the four dissipation combinations. Parameters are $\kappa/\omega=\gamma/\omega=0.5,\delta/\omega=0.05$, $N_\text{{cutoff}} = 200$}.
    \label{fig:Liou_gap}
\end{figure*}

\begin{table*}[t]
\caption{Scaling of the Liouvillian gap $\Delta_L$ with the frequency ratio $\eta$ in the four dissipation combinations, extracted from the finite-size scaling fit $\Delta_L(\eta) = a + b/\eta + c/\eta^2$ at representative points in the SP.}
\label{tab:scaling}
\begin{ruledtabular}
\begin{tabular}{ccccc}
Region & Dissipation & $a$ & $b$ & $c$ \\
\hline
SP & $\kappa$ & $2.6728 \times 10^{-6}$ & $1.3903 \times 10^{-2}$ & $1.5392 \times 10^{-1}$ \\
& $\kappa+\gamma$ & $4.9646 \times 10^{-2}$ & $-1.0086 \times 10^{-2}$ & $-1.3167$ \\
& $\kappa+\delta$ & $6.6503 \times 10^{-2}$ & $1.0818 \times 10^{-3}$ & $-5.4856 \times 10^{-1}$ \\
& $\kappa+\gamma+\delta$ & $1.2375 \times 10^{-1}$ & $6.3176 \times 10^{-2}$ & $-5.6795$ \\
\end{tabular}
\end{ruledtabular}
\end{table*}
\subsection{Principal Component Analysis}
\label{sec:principal}
In the steady state the probability of finding the system in a region of phase space is the fraction of time it spends there. We write the steady state as a mixture of pure states, $\hat\rho_{\mathrm{ss}}=\sum_i p_i|\psi_i\rangle\langle\psi_i|$, and, following Ref.~\cite{xiao2026metastability}, characterize each $|\psi_i\rangle$ by the joint distribution of the spin $s=\pm1$ ($\hat\sigma_z$) and the cavity quadrature $x=(\hat a+\hat a^\dagger)/\sqrt2$,
\begin{equation}
    P_i(s,x)=|\langle s,x|\psi_i\rangle|^2 .
\end{equation}
The similarity of two eigenstates is quantified by the fidelity $F_{ij}=\bigl(\sum_{s,x}\sqrt{P_i(s,x)P_j(s,x)}\bigr)^2$. We connect two eigenstates by an edge of weight $F_{ij}$ whenever $F_{ij}>0.3$, and group them with the Louvain community-detection algorithm. Each group is a principal component of $\hat\rho_{\mathrm{ss}}$, with weight $\sum_{i\in g}p_i$. In the superradiant phase this procedure gives two groups. The first is a spiral channel in which the spin is mostly up, with weight $\operatorname{Tr}(\hat\rho_{\mathrm{spiral}})$ and spin-up population $\langle\hat\sigma_+\hat\sigma_-\rangle_{\mathrm{spiral}}$. The second contains all other states (the two symmetry-broken states and the sigmoid channel), with weight $\operatorname{Tr}(\hat\rho_{\mathrm{rest}})$.
 
We use this to estimate the lifetime of the superradiant state for $\kappa+\gamma$ at $\Omega/\omega=200$, $(\tilde\lambda_x,\tilde\lambda_y)=(1.5,0.5)$ and $\kappa/\omega=\gamma/\omega=0.5$. We assume that the system leaves the rest through the spiral channel, and leaves the spiral channel through a spin jump, which for the dissipator in Eq.~(4) occurs at rate $2\gamma\langle\hat\sigma_+\hat\sigma_-\rangle$. In the steady state, the probability flowing into the spiral channel equals the flow out of it, $2\gamma\langle\hat\sigma_+\hat\sigma_-\rangle_{\mathrm{spiral}}\operatorname{Tr}(\hat\rho_{\mathrm{spiral}})$. The average time spent in the rest before entering the spiral channel is the weight of the rest divided by this flow:
\begin{equation}
\begin{split}
    \tau_{\mathrm{}}
    &\simeq
    \frac{\operatorname{Tr}(\hat\rho_{\mathrm{rest}})}
         {2\gamma\,\langle\hat\sigma_+\hat\sigma_-\rangle_{\mathrm{spiral}}\,
          \operatorname{Tr}(\hat\rho_{\mathrm{spiral}})} \\
    &= \frac{0.953}{2\times0.5\times0.89\times0.047}
    \approx 22.74\,\omega^{-1}.
\end{split}
\end{equation}
This is the lifetime of the symmetry-broken states because every excursion through the spiral channel ends in either symmetry-broken state with probability $1/2$. The rate of switching from one symmetry-broken state to the other is therefore $1/(2\tau_{\mathrm{}})$ in each direction, and the slowest decay rate of this two-state process is the sum of the two, $1/\tau_{\mathrm{}}$.
 
 When atomic dephasing is added ($\kappa+\gamma+\delta$ with $\delta/\omega=0.05$) at the same parameters, the spiral weight doubles, from $0.047$ to $0.094$, so the system enters the spiral channel more often, while the spin-up population of the spiral channel stays almost unchanged, $\langle\hat\sigma_+\hat\sigma_-\rangle_{\mathrm{spiral}}=0.89$. The same flux-balance argument gives
\begin{equation}
    \tau_{\mathrm{}}
    \simeq
    \frac{0.906}{2\times0.5\times0.89\times0.094}
    \approx 10.8\,\omega^{-1}.
\end{equation}
This shows that dephasing shortens the lifetime, by a factor $2.1$. 
The estimate assumes that the spiral channel is left only through spin jumps. Dephasing opens additional exit routes, which increase the outflow from the spiral channel and therefore the rate of entering it.
 
\section{Liouvillian Spectral Gap and Lifetime of Metastable State}
\label{sec:liouvillian}
The principal component analysis (Sec.~\ref{sec:principal}) suggests that atomic dissipation changes how fast the system relaxes. To study this numerically, we use the spectrum of the Liouvillian $\mathcal{L}$ (Eq.~\eqref{eq:master}). The eigenvalues $\lambda_k$ are ordered by decreasing real part. The eigenvalue $\lambda_0=0$ belongs to the steady state, and all other eigenvalues have a negative real part. Each of them describes a mode that decays in time as $e^{-\Gamma_k t}$, where $\Gamma_k=-\mathrm{Re}\,\lambda_k$ is its decay rate. The imaginary part only makes the mode oscillate.

The slowest decay rate $\Gamma_1$ is the Liouvillian gap~\cite{mori2020resolving, mori2023symmetrized}, and $1/\Gamma_1$ is the longest relaxation time of the system. The second rate $\Gamma_2$ belongs to the next eigenvalue.

The Liouvillian gap alone is not enough to decide whether a state is metastable.
We ask what happens to the gap as $\eta=\Omega/\omega$ grows. If $\Gamma_1\to0$, the slow mode never decays and becomes a steady state. This is a genuine dissipative phase transition. If $\Gamma_1$ stays finite, the slow mode has a finite lifetime.
Second, a finite lifetime is called metastable only if it is much longer than the lifetimes of all other modes. After a time of about $1/\Gamma_2$, all faster modes have died out, and only the slowest mode is left. If $\Gamma_2\gg\Gamma_1$, the system then stays in this long-lived state for a long time before it decays at the slow rate $\Gamma_1$~\cite{macieszczak2015towards}. If $\Gamma_2$ is only a little larger than $\Gamma_1$, there is no such window, and the slow mode is just one of several relaxation processes. The ratio $\Gamma_2/\Gamma_1$ measures how clear this separation is, and it is always at least $1$.



\subsection{Cavity photon loss($\kappa$)}
With cavity decay alone, $\Gamma_1$ decreases roughly as $1/\eta$ [Fig.~\ref{fig:Liou_gap}(a)]. Since its rate goes to zero, the two states become degenerate steady states for $\eta\to\infty$, which is a genuine dissipative phase transition~\cite{xiao2026metastability}.

\subsection{Cavity photon loss and atomic relaxation ($\kappa+\gamma$)}
Adding atomic relaxation changes the Liouvillian gap. $\Gamma_1$ no longer decreases with $\eta$. It stays at about $0.05\,\omega$ for all $\eta$ [Fig.~\ref{fig:Liou_gap}(a)], and at $\eta=200$ it is $0.0495\,\omega$, about $650$ times larger than for cavity photon loss alone. The slowest mode is still parity-odd, so it still describes the jumps between the two symmetry-broken states, but now these jumps happen at a finite rate. The ratio $\Gamma_2/\Gamma_1\approx10$, and it stays close to $10$ for all $\eta$ [Fig.~\ref{fig:Liou_gap}(b)].

So the gap is finite and well separated from the other rates. The superradiant state is therefore metastable, with a lifetime $1/\Gamma_1\approx20.142\,\omega^{-1}$. Atomic relaxation returns the atom to its ground state, and this lets the system move between the two symmetry-broken states. This is the same process that appears as the bridge between the two peaks in the Wigner function.

\subsection{Cavity photon loss and atomic dephasing ($\kappa+\delta$)}
Adding atomic dephasing to cavity photon loss now changes the Liouvillian gap. It does not close as the $\eta$ increases but rather settles at $\Gamma_1 = 0.065 \omega$. But the ratio $\Gamma_2/\Gamma_1 \approx 1$. Therefore, there is no seperation of timescales between the decay modes, and no metastable symmetry-broken state.



\subsection{Cavity photon loss, atomic relaxation and dephasing ($\kappa+\gamma+\delta$)}

Adding atomic dephasing to $\kappa+\gamma$ makes the decay faster. For $\delta/\omega=0.05$ the gap increases to $\Gamma_1\simeq0.124\,\omega$, so the lifetime $1/\Gamma_1$ drops to $8.1\,\omega^{-1}$ at $\eta=200$. The gap is finite for all $\eta$ [Fig.~\ref{fig:Liou_gap}(a), Table~\ref{tab:scaling}]. The second rate is $\Gamma_2\simeq0.51\,\omega$, so $\Gamma_2/\Gamma_1\approx4.1$[Fig.~\ref{fig:Liou_gap}(b)]. The gap is finite and still separated from the rest of the spectrum, so the metastability is present, but its lifetime is reduced. The slowest mode remains parity-odd.
 

Therefore, Cavity photon loss alone gives a genuine phase transition, because $\Gamma_1\to0$. Adding atomic relaxation to cavity photon loss makes $\Gamma_1$ finite while keeping $\Gamma_2/\Gamma_1\approx10$, so the superradiant state becomes metastable with a lifetime of about $20.14\,\omega^{-1}$. Adding atomic dephasing to cavity photon loss and atomic relaxation then shortens this lifetime to $8.08\,\omega^{-1}$  and reduces the ratio $\Gamma_2/\Gamma_1\approx 4.1$. 

\section{Conclusion}
\label{sec:conclusion}
We have investigated the nonequilibrium dynamics of the anisotropic open Rabi model in the presence of cavity photon loss, atomic relaxation, and atomic dephasing. By combining semiclassical stability analysis with exact Liouvillian spectral calculations, phase-space tomography through Wigner functions, and principal component analysis, we have demonstrated that different dissipation channels can lead to qualitatively distinct forms of dissipative criticality and metastable dynamics.

Our mean-field analysis reveals that atomic relaxation and atomic dephasing have fundamentally different effects on the superradiant phase. While atomic relaxation preserves the qualitative structure of the phase diagram, atomic dephasing in the absence of relaxation destabilizes the superradiant fixed points, leaving only the normal phase stable. This distinction originates from their different microscopic actions: atomic relaxation modifies the population dynamics while atomic dephasing suppresses the coherence required to sustain superradiant order.

The phase-space analysis further illustrates how these competing dissipation mechanisms reshape the steady-state structure. In the presence of cavity photon loss alone, the Wigner distribution exhibits two well-separated peaks associated with the symmetry-broken superradiant configurations. Adding atomic relaxation generates an interference-textured connection between these peaks, indicating dynamical excursions between the two configurations. In contrast, atomic dephasing suppresses the superradiant peaks and redistributes the phase-space weight towards the normal-phase configuration. When both atomic channels are present, the resulting steady state becomes strongly delocalized, reflecting the competition between population relaxation and coherence loss.

The Liouvillian spectral analysis provides a complementary characterization of these dynamical regimes. With cavity photon loss alone, the Liouvillian gap closes approximately as \(1/\eta\), where \(\eta=\Omega/\omega\), indicating the emergence of degenerate symmetry-broken steady states in the large-frequency-ratio limit and thus a genuine dissipative phase transition. The inclusion of atomic relaxation arrests this gap closing and produces a finite but well-separated slow decay mode, establishing the superradiant state as metastable rather than asymptotically stable. For the parameters considered, the associated lifetime is approximately \(20\,\omega^{-1}\). Adding atomic dephasing to this configuration increases the Liouvillian gap, reducing the metastable lifetime to approximately \(8\,\omega^{-1}\), while the separation between the slowest decay modes also becomes less pronounced. In contrast, when dephasing is added to cavity photon loss without atomic relaxation, the gap remains finite but the ratio of the two slowest decay rates approaches unity, indicating the absence of a well-isolated metastable superradiant mode.

Taken together, these results establish that the microscopic nature of environmental coupling plays a crucial role in determining the fate of superradiant order in finite-component light–matter systems. In particular, dissipation need not simply suppress coherent dynamics: depending on the underlying relaxation and dephasing processes, it can stabilize long-lived symmetry-broken configurations, generate transitions between them, or completely suppress their existence. Our findings highlight the importance of distinguishing different environmental channels when characterizing dissipative phase transitions and metastability, and provide a framework for exploring their interplay in experimentally accessible cavity and circuit QED systems.

\section*{Appendix A: Numerical determination of semiclassical fixed points}
\label{app:meanfield_numerics}

The semiclassical phase diagram is obtained from the fixed points of
the five-dimensional nonlinear dynamical system
Eqs.~\eqref{eq:x}--\eqref{eq:z}. It is convenient to collect the real
dynamical variables into the vector
\begin{equation}
    \mathbf{s}=(x,y,u,v,z),
\end{equation}
such that the mean-field equations can be written compactly as
\begin{equation}
    \dot{\mathbf{s}}
    =
    \mathbf{F}
    \left(
    \mathbf{s};
    \tilde{\lambda}_x,\tilde{\lambda}_y
    \right).
    \label{eq:app_vector_field}
\end{equation}
A semiclassical steady state is therefore a fixed point
$\mathbf{s}_*$ satisfying
\begin{equation}
    \mathbf{F}
    \left(
    \mathbf{s}_*;
    \tilde{\lambda}_x,\tilde{\lambda}_y
    \right)
    =\mathbf{0}.
    \label{eq:app_fixedpoint}
\end{equation}
In the presence of atomic dissipation, the Bloch-sphere equality $4(u^2+v^2)+z^2=1$ is generally not preserved. Consequently, the superradiant fixed points cannot in general be reduced to a closed analytical expression and must be determined from the full set of five nonlinear equations. We therefore use numerical root finding together with analytic continuation in parameter space to determine the nontrivial branches and subsequently assess their linear stability.

\subsection*{A.1 Normal-phase fixed point}
The normal phase (NP) is an exact fixed point of the semiclassical
equations and is given by
\begin{equation}
    \mathbf{s}_{\rm NP}
    =
    (0,0,0,0,-1).
    \label{eq:app_np}
\end{equation}
At this point the cavity amplitude and transverse atomic coherence vanish, while the atom remains in its ground state. The stability of this fixed point can be determined analytically from the Jacobian of the mean-field equations. As shown in Sec.~\ref{subsec:NP}, the longitudinal-spin fluctuation decouples from the remaining four variables, and the NP stability boundary is given by Eq.~\eqref{eq:NP_boundary}. This analytical condition is used directly in constructing the semiclassical phase diagram.

\subsection*{A.2 Superradiant fixed points and their stability}
\label{app:sp_numerics}
\begin{figure*}[t]
    \includegraphics[width=0.88\textwidth]{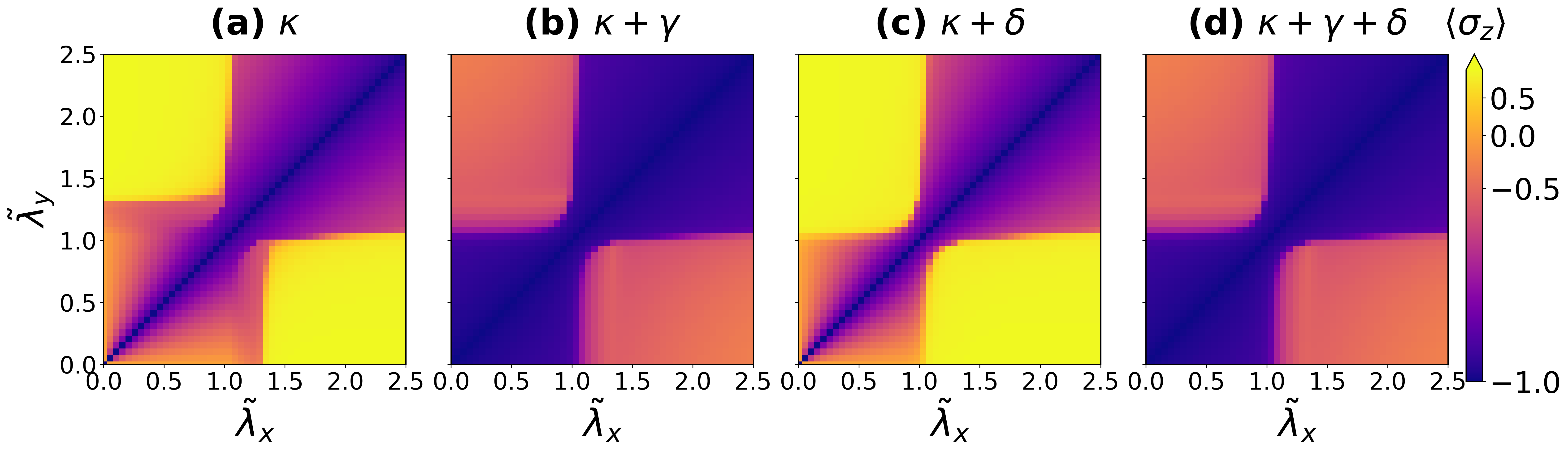}
    \caption{Steady-state \(\langle \sigma_z \rangle\) obtained from exact Liouvillian diagonalization, plotted over the \((\tilde{\lambda}_x, \tilde{\lambda}_y)\) plane for different combinations of cavity decay, atomic decay, and atomic dephasing. The parameters taken here are $\omega = 1$, $\Omega = 1000$ $\kappa = 0.5$, $\gamma = 0.2, \delta = 0.01$. The bosonic cutoff is: $N_{\text{cutoff}}=300$.}
    \label{fig:steady_state_sigmaz}
\end{figure*}
The superradiant phase (SP) corresponds to nontrivial fixed points with
\begin{equation}
    \alpha_* = x_*+iy_* \neq 0.
\end{equation}
For finite atomic dissipation, the five equations
\begin{equation}
    \dot{x}=\dot{y}=\dot{u}=\dot{v}=\dot{z}=0
\end{equation}
form a coupled nonlinear system. We solve this system numerically at each point in the $(\tilde{\lambda}_x,\tilde{\lambda}_y)$ parameter plane.

The numerical procedure is based on a multidimensional root-finding algorithm. To improve convergence and to follow continuous superradiant branches, the solution obtained at a neighboring point in parameter space is used as an initial guess for the next point. This continuation procedure is supplemented by additional initial guesses obtained from small perturbations of the converged solution. Both forward and reverse parameter scans are performed to reduce the possibility of missing a stable branch because of the choice of initial condition.

For a candidate fixed point $\mathbf{s}_*$, convergence is assessed from the residual
\begin{equation}
    R(\mathbf{s}_*)
    =
    \left\|
    \mathbf{F}
    \left(
    \mathbf{s}_*;
    \tilde{\lambda}_x,\tilde{\lambda}_y
    \right)
    \right\|.
    \label{eq:app_residual}
\end{equation}
Only solutions with a residual below the prescribed numerical tolerance are retained. To distinguish nontrivial superradiant solutions from the normal-phase solution, we additionally require a finite transverse amplitude,
\begin{equation}
    r_{\perp}
    =
    \sqrt{x_*^2+y_*^2+u_*^2+v_*^2},
    \label{eq:app_rperp}
\end{equation}
above a small numerical threshold. The longitudinal spin is restricted to its physical range,
\begin{equation}
    -1\leq z_*\leq 1.
\end{equation}
In the parameter regimes considered in this work, the converged superradiant solutions satisfy $z_*\leq0$.

The $\mathbb{Z}_2$ parity symmetry of the model implies that every nonzero fixed point has a symmetry-related partner,
\begin{equation}
    (x_*,y_*,u_*,v_*,z_*)
    \longleftrightarrow
    (-x_*,-y_*,-u_*,-v_*,z_*).
    \label{eq:app_z2}
\end{equation}
The two branches have identical linear stability spectra. Therefore, only one representative of each symmetry-related pair is required when constructing the phase diagram. Once a nontrivial fixed point has been obtained, its stability is determined from the Jacobian
\begin{equation}
    J_{ij}=\left.\frac{\partial F_i}{\partial s_j}\right|_{\mathbf{s}=\mathbf{s}_*}.
    \label{eq:app_jacobian}
\end{equation}
For the variables $(x,y,u,v,z)$, the Jacobian is presented in Eqn.~\eqref{eq:jac_sp} in the main text.

\subsection*{A.3 Identification of the bistable region}
\label{app:bistability}
The semiclassical bistable region is identified by comparing the stability of the normal and superradiant fixed points at the same point in parameter space. The NP is locally stable when the analytical condition associated with Eq.~\eqref{eq:NP_boundary} is satisfied, whereas the SP is locally stable when $\mathrm{Re}(\mu_i) < 0$ is satisfied.

The bistable region is therefore defined by the simultaneous conditions
\begin{equation}
    \text{NP stable}
    \qquad\text{and}\qquad
    \text{SP stable}.
    \label{eq:app_bistability}
\end{equation}
Equivalently, if $\mu_i^{\rm SP}$ denote the eigenvalues of the SP Jacobian, the condition can be written as
\begin{equation}
    \max_i\operatorname{Re}\mu_i^{\rm SP}<0
\end{equation}
together with the NP stability condition.

\section*{Appendix B: Longitudinal-spin order parameter}

In addition to the photon number, the
longitudinal collective spin \(\langle s_z \rangle\) provides a complementary
order parameter that tracks the atomic inversion across the normal,
superradiant, and bistable regimes (see Fig.~\ref{fig:steady_state_sigmaz}). For each point in the
\((\tilde{\lambda}_x,\tilde{\lambda}_y)\)-plane we evaluate \(\langle s_z \rangle\) in the steady
state. These plots show how the atoms evolve from nearly fully
inverted configurations with \(\langle s_z \rangle \approx -1\) in the normal
phase to partially excited states in the superradiant phase. The behaviour of \(\langle s_z \rangle\) thus offers a clear,
spin-based visualization of the phase boundaries.

\section*{Appendix C: Liouvillian spectral decomposition}
\label{app:liouvillian_spectrum}
The full quantum dynamics of the cavity--spin system is governed by the Lindblad master equation
\begin{equation}
    \frac{d\rho}{dt}
    =
    \mathcal{L}\rho,
    \label{eq:app_master}
\end{equation}
where $\mathcal{L}$ is the Liouvillian superoperator defined in Eq.~(\ref{eq:liouvillian}). Because $\mathcal{L}$ is generally non-Hermitian, its spectral decomposition requires separate left and right eigenoperators.

The right eigenoperators are defined by
\begin{equation}
    \mathcal{L}R_j
    =
    \lambda_j R_j,
    \label{eq:app_right_eigen}
\end{equation}
while the corresponding left eigenoperators satisfy
\begin{equation}
    L_j^\dagger\mathcal{L}
    =
    \lambda_j L_j^\dagger.
    \label{eq:app_left_eigen}
\end{equation}
Here $\lambda_j$ are, in general, complex Liouvillian eigenvalues.

For a trace-preserving Lindblad evolution, the stationary state corresponds to an eigenvalue
\begin{equation}
    \lambda_0=0,
\end{equation}
with
\begin{equation}
    \mathcal{L}[\rho_{\mathrm{ss}}]=0.
    \label{eq:app_stationary}
\end{equation}
The remaining eigenvalues determine the decay and oscillation timescales of perturbations around the stationary state. For a stable finite system,
\begin{equation}
    \operatorname{Re}\lambda_j<0,
    \qquad j\neq0.
\end{equation}

When the Liouvillian is diagonalizable, the left and right eigenoperators can be chosen to satisfy the biorthonormality condition
\begin{equation}
    \operatorname{Tr}
    \left[
        L_i^\dagger R_j
    \right]
    =
    \delta_{ij}.
    \label{eq:app_biorthogonal}
\end{equation}
An arbitrary initial density matrix can then be expanded as
\begin{equation}
    \rho(0)
    =
    \sum_j c_j R_j,
    \qquad
    c_j=
    \operatorname{Tr}
    \left[
        L_j^\dagger\rho(0)
    \right],
    \label{eq:app_initial_expansion}
\end{equation}
and its subsequent evolution is
\begin{equation}
    \rho(t)
    =
    \sum_j
    c_j e^{\lambda_j t}R_j.
    \label{eq:app_time_evolution}
\end{equation}

Eq.~\eqref{eq:app_time_evolution} makes clear why eigenvalues whose real parts are close to zero are particularly important. Such modes decay slowly and can dominate the dynamics over an extended intermediate-time window before the system reaches the stationary state. These slowly decaying modes provide the spectral description of metastable dynamics.

The steady state itself is the zero-eigenvalue right eigenoperator $R_0$, normalized according to
\begin{equation}
    \operatorname{Tr}\rho_{\mathrm{ss}}=1.
\end{equation}
Thus, the nonzero Liouvillian eigenmodes should not be interpreted as additional stationary density matrices. Rather, they describe relaxation channels toward the unique steady state and, when their decay rates are anomalously small, the associated metastable dynamics.
\begin{figure*}[t]
    \centering
    \includegraphics[width=\textwidth]{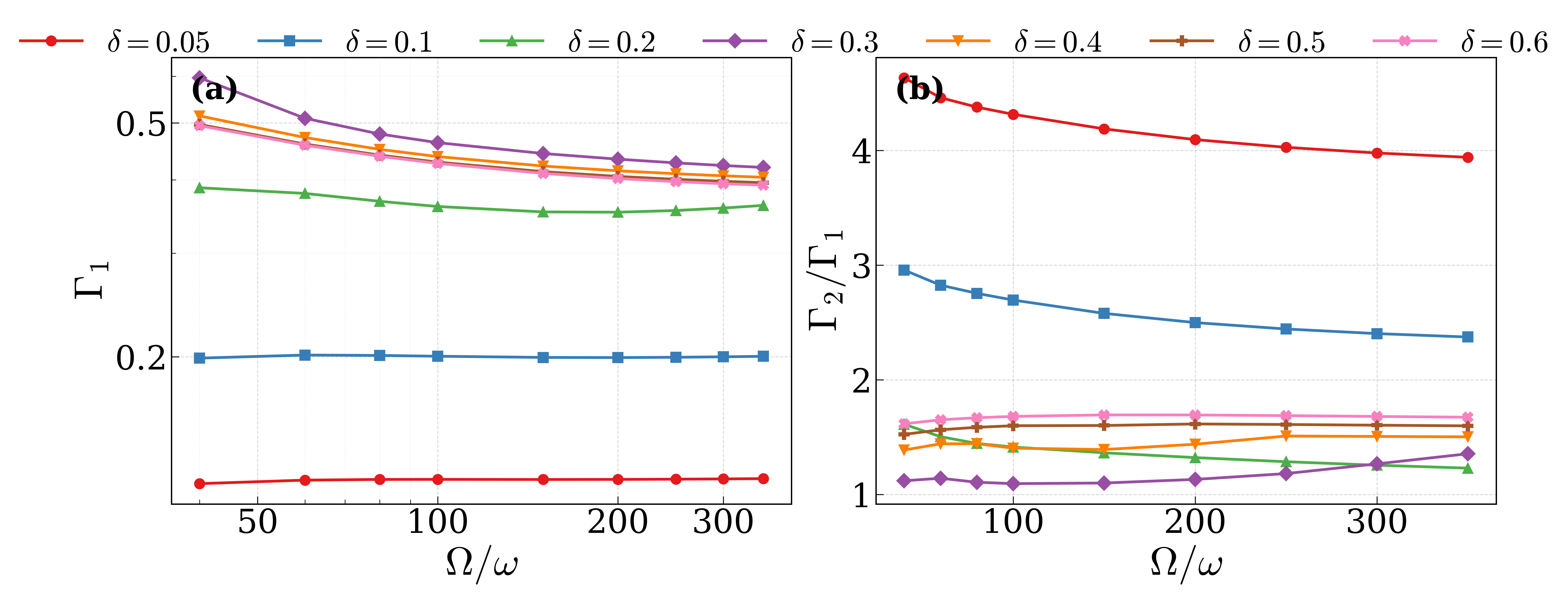}
    \caption{(a) Liouvillian gap $\Gamma_1$ and (b) ratio $\Gamma_2/\Gamma_1$ versus $\eta=\Omega/\omega$ for $\kappa+\gamma+\delta$ at several values of the dephasing rate $\delta/\omega$. The remaining parameters are $(\tilde\lambda_x,\tilde\lambda_y)=(1.5,0.5)$ and $\kappa/\omega=\gamma/\omega=0.5$.}
    \label{fig:delta_scan}
\end{figure*}
\section*{Appendix D: Steady-state calculation in Liouville space}
\label{app:liouville_space}
To obtain the exact quantum steady state numerically, we truncate the bosonic Hilbert space at a maximum photon number $N_{\mathrm{cutoff}}$. Since the atomic Hilbert space is two dimensional, the total Hilbert space dimension is
\begin{equation}
    d=2(N_{\mathrm{cutoff}}+1).
    \label{eq:app_hilbert_dimension}
\end{equation}

We use the product basis
\begin{equation}
    \{|s,n\rangle\},
    \qquad
    s\in\{\uparrow,\downarrow\},
    \qquad
    n=0,\ldots,N_{\mathrm{cutoff}}.
\end{equation}
The density matrix is written as
\begin{equation}
    \rho
    =
    \sum_{\alpha,\beta}
    \rho_{\alpha\beta}
    |\alpha\rangle\langle\beta|,
    \label{eq:app_density_expansion}
\end{equation}
where $\alpha$ and $\beta$ collectively denote the atom and photon indices.

The density matrix is then vectorized according to a fixed vectorization convention,
\begin{equation}
    |\rho\rangle\rangle
    =
    {\rm vec}(\rho),
    \label{eq:app_vectorization}
\end{equation}
so that the master equation becomes an ordinary linear differential equation in Liouville space,
\begin{equation}
    \frac{d}{dt}|\rho\rangle\rangle
    =
    \mathbf{L}|\rho\rangle\rangle.
    \label{eq:app_liouville_matrix}
\end{equation}
The Liouvillian matrix $\mathbf L$ therefore has dimension
\begin{equation}
    d_{\mathcal L}
    =
    d^2
    =
    4(N_{\mathrm{cutoff}}+1)^2.
    \label{eq:app_liouvillian_dimension}
\end{equation}

For completeness, using column-stacking vectorization,
\begin{equation}
    {\rm vec}(A\rho B)
    =
    (B^{\mathsf T}\otimes A)
    {\rm vec}(\rho),
\end{equation}
the coherent part of the Liouvillian is represented by
\begin{equation}
    \mathbf L_H
    =
    -i
    \left(
        \mathbbm{1}\otimes H
        -
        H^{\mathsf T}\otimes\mathbbm{1}
    \right).
    \label{eq:app_coherent_liouvillian}
\end{equation}
For the dissipator convention used in Eq.~\eqref{eq:dissipator}, the contribution of a jump operator $L_\mu$ is
\begin{equation}
    \mathbf L_\mu
    =
    2L_\mu^*\otimes L_\mu
    -
    \mathbbm{1}\otimes L_\mu^\dagger L_\mu
    -
    (L_\mu^\dagger L_\mu)^{\mathsf T}
    \otimes\mathbbm{1}.
    \label{eq:app_dissipative_liouvillian}
\end{equation}
Consequently, the full matrix representation is
\begin{equation}
    \mathbf L
    =
    \mathbf L_H
    +
    \kappa\,\mathbf L_a
    +
    \gamma\,\mathbf L_{\sigma_-}
    +
    \delta\,\mathbf L_{\sigma_z}.
    \label{eq:app_full_liouvillian}
\end{equation}

The stationary state is obtained from the null-space equation
\begin{equation}
    \mathbf L|\rho_{\mathrm{ss}}\rangle\rangle
    =
    0,
    \label{eq:app_nullspace}
\end{equation}
together with the normalization condition $\operatorname{Tr}\rho_{\mathrm{ss}}=1$.
Equivalently, the steady state can be obtained as the right eigenvector corresponding to the Liouvillian eigenvalue $\lambda_0=0$, followed by normalization to unit trace.

In practice, the same truncated Hilbert-space representation is used for calculating the steady-state observables and the Liouvillian spectrum. The steady-state photon number and spin observables are obtained from
\begin{equation}
    \langle\hat O\rangle_{\mathrm{ss}}
    =
    \operatorname{Tr}
    \left(
        \rho_{\mathrm{ss}}\hat O
    \right),
    \label{eq:app_observables}
\end{equation}
with,
\begin{equation}
    n_{\mathrm{ph}}
    =
    \langle a^\dagger a\rangle_{\mathrm{ss}},
    \qquad
    S_z
    =
    \langle\sigma_z\rangle_{\mathrm{ss}}.
    \label{eq:app_order_parameters}
\end{equation}

\section*{Appendix E: Dependence of the Liouvillian gap on the dephasing rate}
\label{app:delta}

In Sec.~\ref{sec:liouvillian} we fixed the dephasing rate at $\delta/\omega=0.05$. Here we show how the Liouvillian gap $\Gamma_1$ and the ratio $\Gamma_2/\Gamma_1$ change when $\delta$ is varied for the case ($\kappa+\gamma+\delta$). Fig.~\ref{fig:delta_scan} shows both quantities as functions of $\eta=\Omega/\omega$ for $\delta/\omega=0.05,\,0.1,\,0.2,\,0.3,\,0.4,\,0.5$ and $0.6$, at $(\tilde\lambda_x,\tilde\lambda_y)=(1.5,0.5)$ and $\kappa/\omega=\gamma/\omega=0.5$.

For every value of $\delta$ the gap $\Gamma_1$ stays finite and approaches a constant as $\eta$ increases [Fig.~\ref{fig:delta_scan}(a)]. Dephasing therefore never restores a closing gap in the presence of atomic relaxation, and the symmetry-broken states always have a finite lifetime $1/\Gamma_1$. The size of the gap depends on $\delta$. It grows from $\Gamma_1\approx0.12\,\omega$ at $\delta/\omega=0.05$ to $\Gamma_1\approx0.2\,\omega$ at $\delta/\omega=0.1$, and so on. Hence dephasing shortens the lifetime of the superradiant state most strongly at small $\delta$.

The ratio $\Gamma_2/\Gamma_1$ decreases quickly with $\delta$ [Fig.~\ref{fig:delta_scan}(b)]. It falls from $\Gamma_2/\Gamma_1\approx4$ at $\delta/\omega=0.05$ to about $2.5$ at $\delta/\omega=0.1$. For $\delta/\omega\gtrsim0.2$ it lies between $1$ and $1.7$ for all $\eta$. In this range, $\Gamma_1$ is no longer separated from the rest of the spectrum, and the slow mode is one of several comparable relaxation processes, and the metastability is lost. This agrees with the $\kappa+\delta$ case discussed in Sec.~\ref{sec:liouvillian}, where $\Gamma_2/\Gamma_1\approx1$.





\bibliographystyle{apsrev4-2}
\bibliography{ref}

\end{document}